\documentclass{PoS}

\usepackage{amsmath,amssymb,bm}
\usepackage{array}
\newcommand{\tvec}[1]{\boldsymbol{#1}}

\allowdisplaybreaks

\title{Framework for evolution in double parton scattering}

\ShortTitle{Framework for evolution in double parton scattering}

\author{\speaker{M.G.A. Buffing}
	\thanks{The work presented here has been performed in collaboration with Markus Diehl and Tomas Kasemets.}\\
	Deutsches Elektronen-Synchrotron DESY\\
	22603 Hamburg, Germany\\
	E-mail: \email{maarten.buffing@desy.de}}

\abstract{Double parton scattering (DPS) describes two colliding hadrons having interactions in the form of two hard processes, each initiated by a separate pair of partons. Just as for single parton scattering, the resummation of soft gluon exchange gives rise to a soft function, which is a necessary ingredient for obtaining rapidity evolution equations. For various regions of phase space, we derive the rapidity evolution and the scale evolution of double transverse momentum dependent parton distribution functions (DTMDs) as well as of the $p_{T}$-resummed cross section for double Drell-Yan like processes. This contributes to a framework that can be used for phenomenological DPS studies including resummation.}

\FullConference{XXV International Workshop on Deep-Inelastic Scattering and Related Subjects\\
		3-7 April 2017\\
		University of Birmingham, UK}

\begin{document}
%%%%%%%%%%%%%%%%%%%%%%%%%%%%%%%%%%%%%%%%%%%%%%%%%%%%%%%%%%%%%%%%%%%%%%%%%%%%%%%

%%%%%%%%%%%%%%%%%%%%%%%%%%%%%%%%%%%%%%%%%%%%%%%%%%%%%%%%%%%%%%%%%%%%%%%%%%%%%%%
\section{Introduction}
\label{ss:introduction}
\begin{figure}[!b]
\begin{center}
\includegraphics[width=0.5\textwidth]{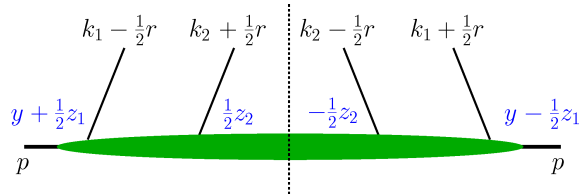}
%Width = 1941, height = 650
\caption{Illustration of the momenta $k_{i}$ and $r$ as well as the impact parameter space variables $\textcolor{blue}{z_{i}}$ and $\textcolor{blue}{y}$ (blue color online) that are involved in double parton scattering.}
\label{f:labeling}
\end{center}
\end{figure}

%%%%%%%%%%%%%%%%%%%%%%%%%%%%%%%%%%%%%%%%%%%%%%%%%%%%%%%%%%%%%%%%%%%%%%%%%%%%%%%
In double parton scattering (DPS), firstly described in the Refs.~\cite{Landshoff:1975eb,Landshoff:1978fq}, two hard partonic processes take place in a single hadron-hadron collision and it is as such relevant for the LHC~\cite{Bartalini:2011jp,Proceedings:2016tff}. Just as for single parton scattering (SPS), one needs parton distribution functions (PDFs) for describing the nonperturbative dynamics in the initial state. Specifically, one needs double PDFs (DPDFs) and double transverse momentum dependent PDFs (DTMDs). In Fig.~\ref{f:labeling} the relevant momentum and configuration space variables that are required to describe DTMDs are illustrated. Not only does one need the positions $\tvec{z}_1$ and $\tvec{z}_2$, the distance $\tvec{y}$ between the two hard processes in configuration space has to be introduced as well.

In these proceedings, we limit ourselves to color singlet production processes such as double Drell-Yan or the production of two Higgs bosons. There are different regions of phase-space that have to be taken into consideration. We treat all of them, see also \cite{Buffing:2017}, but here we focus on perturbative $|\tvec{z}_1|,\,|\tvec{z}_2| \ll |\tvec{y}|$, with $\tvec{y}$ fixed. This implies that we are looking at two perturbative hard partonic interactions that are separated from each other in configuration space~\cite{Buffing:2017,Buffing:2016qql}.

We generalize the resummation formalism for the single Drell-Yan process in Ref.~\cite{Collins:1984kg} to double Drell-Yan. In particular, we focus on the rapidity and energy scale evolution equations and their solutions. For SPS, the procedure for TMD evolution is known~\cite{Aybat:2011zv}. We show how these results can be generalized for the double color singlet production in DPS, see also~\cite{Buffing:2017,Buffing:2016qql}. On top, we also give the matching equations for DTMD/DPDF matching.

%%%%%%%%%%%%%%%%%%%%%%%%%%%%%%%%%%%%%%%%%%%%%%%%%%%%%%%%%%%%%%%%%%%%%%%%%%%%%%%
\section{Evolution}
\label{ss:evolution}
%%%%%%%%%%%%%%%%%%%%%%%%%%%%%%%%%%%%%%%%%%%%%%%%%%%%%%%%%%%%%%%%%%%%%%%%%%%%%%%
Just as for single Drell-Yan like processes, describing cross sections and DTMDs in DPS involves the handling of a soft function, albeit a more complicated one, since it involves two partonic processes simultaneously. Since the energy scale evolution is related to the soft function through the rapidity evolution kernel $K$, we need to ensure a proper handling of the soft function in order to get a well-defined starting point for the evolution equations. In these proceedings we will give the results for the soft function and refer to \cite{Buffing:2017} for details. For other literature see e.g.\ Ref.~\cite{Collins:2011zzd} for a general discussion and Ref.~\cite{Vladimirov:2016qkd} for the DPS soft factor at two loops.

It is necessary to make a few remarks about the color structure first. For SPS, Drell-Yan only has one possible color structure for the two partons in the t-channel, namely a color singlet. For DPS, the situation is more involved, since there are more ways to connect the four initial state (t-channel) parton lines to (s-channel) color singlet final states~\cite{Kasemets:2014yna}. Therefore, for a DPS situation involving (anti)quarks only, we have a color singlet and octet configuration, whereas for gluons (or mixed states) the situation becomes even more complicated. As such, we introduce the parameters $R$ and $R'$ to label the color configuration. In DPS we need two such parameters due to having two partonic processes.

In the short distance expansion it can then be proven~\cite{Buffing:2017} that the soft function for DPS factorizes as
\begin{align}
{}^{RR'}S_{a_1 a_2}(\tvec{z}_1, \tvec{z}_2, \tvec{y}) & = {}^{R\,}C_{s, a_1}(\tvec{z}_1)\, {}^{R\,}C_{s, a_2}(\tvec{z}_2)\,{}^{RR}S(\tvec{y})\, \delta_{RR'}^{} \, , \label{e:soft-fact-match_B}
\end{align}
implying that the soft function can be factorized in $\tvec{z}_1$, $\tvec{z}_2$ and $\tvec{y}$ dependent parts. Furthermore, the soft function is diagonal in the color representations $R$ and $R'$. Given the above form of the soft function, it can be shown that the corresponding rapidity evolution kernel $K$ is given by
\begin{align}
{}^{RR'}K_{a_1 a_2}(\tvec{z}_i,\tvec{y}; \mu_i) & = \delta_{RR'}^{}\, \bigl[{}^{R}K_{a_1}(\tvec{z}_1;\mu_1) + {}^{R}K_{a_2}(\tvec{z}_2;\mu_2) + {}^{R}{J(\tvec{y}; \mu_i)}\bigr]\, , \label{e:CS-gen-match}
\end{align}
where the additive structure of the different parts is a consequence of the factorized form of the soft function in Eq.~\ref{e:soft-fact-match_B}. In the above equation and all further occurrences, $\mu_i$ is a shorthand notation to indicate both $\mu_1$ and $\mu_2$, which are the renormalization scales for the partons $1$ and $2$. We similarly use the shorthand notation $x_i$ and $\tvec{z}_i$.

The evolution in the rapidity scale parameter $\zeta$ is given by
\begin{align}
\frac{\partial}{\partial\log \zeta} 
{}^{R}{F_{a_1 a_2}}(x_i,\tvec{z}_i,\tvec{y}, \mu_i,\zeta) & = \frac{1}{2} \sum_{R'}{}^{RR'}{K_{a_1 a_2}(\tvec{z}_i,\tvec{y}; \mu_i)}\,{}^{R'}{F_{a_1 a_2}(x_i,\tvec{z}_i,\tvec{y}; \mu_i,\zeta)} \, , \label{e:CS-TMD}
\end{align}
with the evolution kernel given in Eq.~\ref{e:CS-gen-match}. In this equation, the DTMD has absorbed the soft function $S$. The evolution kernel $K$ contains a $\mu$-dependence that has to be understood before writing down the full energy scale evolution equations. See for example Chapter~6 of \cite{Buffing:2017} or Section~3 of \cite{Buffing:2016qql} for a detailed discussion hereof. Solving the differential equation in Eq.~\ref{e:CS-TMD} gives us
\begin{align}
	\label{e:CS-TMD-sol}
{}^{R}{F}_{a_1 a_2}(x_i,\tvec{z}_i,\tvec{y};\mu_i,\zeta)
&= \sum_{R'} {}^{RR'}{\exp}\biggl[
		K_{a_1 a_2}(\tvec{z}_i,\tvec{y};\mu_i)
			\log \frac{\sqrt{\zeta}}{\sqrt{\zeta_0}} \,\biggr]
	{}^{R'}{F}_{a_1 a_2}(x_i,\tvec{z}_i,\tvec{y};\mu_i,\zeta_0) \, ,
\end{align}
where the object ${}^{RR'}{\exp}$ is a matrix exponential defined through
\begin{align}
	\label{e:matrixexponential}
{}^{RR'}{\exp}(M) &= \delta_{RR'} + {}^{RR'}M + \sum_{n=0}^\infty \sum_{R_2,\ldots,R_{n}}
	\frac{{}^{RR_2}{M}\cdots {}^{R_{n} R'}{M}}{n!}\,.
\end{align}

The aforementioned energy evolution of DTMDs is governed by
\begin{align}
\frac{\partial}{\partial \log\mu_1}\,{}^{R}F_{a_1 a_2}(x_i,\tvec{z}_i,\tvec{y};\mu_i,\zeta) & = \gamma_{F, a_1}(\mu_1, x_1\zeta/x_2)\,{}^{R}F_{a_1 a_2}(x_i,\tvec{z}_i,\tvec{y};\mu_i,\zeta) \label{e:RG-TMD}
\end{align}
for $\mu_1$ and through a similar expression for $\mu_2$. In the above, the $\gamma_{F, a_1}(\mu_1, x_1\zeta/x_2)$ is the same anomalous dimension that appears in the evolution of single TMDs as well. For a consistent description, the $\zeta$-argument has to be multiplied by either $x_1/x_2$ or $x_2/x_1$, depending on which partonic process is described. Solving for both the energy scales $\mu_1$ and $\mu_2$ then gives
\begin{align}
{}^{R}F_{a_1 a_2}(x_i,\tvec{z}_i,\tvec{y};\mu_i,\zeta) = & {}^{R}F_{a_1 a_2}(x_i,\tvec{z}_i,\tvec{y};\mu_{0i},\zeta) \nonumber \\[0.2em]
& \qquad \times\exp\biggl[ \int_{\mu_{01}}^{\mu_1} \frac{d\mu}{\mu}\,\gamma_{F,a_1}(\mu, x_1\zeta/x_2) + \int_{\mu_{02}}^{\mu_2} \frac{d\mu}{\mu}\,\gamma_{F,a_2}(\mu, x_2\zeta/x_1) \biggr], \label{e:RG-TMD-sol}
\end{align}
doubling the SPS results. This was to be expected, since the $\mu$ evolution comes from hard loops associated with only one of the two partons, resulting in two independent contributions to the solution of the evolution equations. \newline

Combining the results in the Eqs.~\ref{e:CS-TMD-sol} and \ref{e:RG-TMD-sol}, we get the combined solution
\begin{align}
{}^{R}F_{a_1 a_2}(x_i,\tvec{z}_i,\tvec{y};\mu_i,\zeta) & = \exp\,\biggl\{ \int_{\mu_{01}}^{\mu_1} \frac{d\mu}{\mu}\,\biggl[\gamma_{F,a_1}(\mu, \mu^2) - \gamma_{K,a_1}(\mu) \log\frac{\sqrt{x_1\zeta/x_2}}{\mu} \biggr] \nonumber \\
& \hspace{9.5mm} + \int_{\mu_{02}}^{\mu_2} \frac{d\mu}{\mu}\,\biggl[ \gamma_{F,a_2}(\mu, \mu^2) - \gamma_{K,a_2}(\mu) \log\frac{\sqrt{x_2\zeta/x_1}}{\mu} \biggr] \nonumber \\
& \hspace{9.5mm} + \Bigl[ {}^{R}K_{a_1}(\tvec{z}_1,\mu_{01}) + {}^{R}K_{a_2}(\tvec{z}_2,\mu_{02}) + {}^{R}J(\tvec{y},\mu_{0i}) \Bigr]\log\frac{\sqrt{\zeta}}{\sqrt{\zeta_0}} \biggr\} \nonumber \\
& \quad \times {}^{R}F_{a_1 a_2}(x_i,\tvec{z}_i,\tvec{y};\mu_{0i},\zeta_0) \, , \label{e:evsolving}
\end{align}
where both the rapidity and evolution scale evolution have been taken into account. In this, the $\zeta_0$, $\mu_{01}$ and $\mu_{02}$ are the starting scales of the three evolution parameters. The presence of the anomalous dimensions $\gamma_k$ is a consequence of the $\mu$-dependence of the evolution kernel ${}^{RR'}{K_{a_1 a_2}(\tvec{z}_i,\tvec{y}; \mu_i)}$ in Eq.~\ref{e:CS-TMD}. It satisfies
\begin{align}
\gamma_{K,a}(\mu) &= {}^{R}\gamma_{K,a}(\mu) + {}^{R}\gamma_J (\mu), \label{e:AD-sum}
\end{align}
with ${}^{R}\gamma_{K,a}(\mu)$ and ${}^{R}\gamma_J (\mu)$ given by
\begin{align}
\frac{\partial}{\partial \log\mu_1}\, {}^{R} K_a(\bm{z}; \mu_1) & = - {}^{R}\gamma_{K,a}(\mu_1), \label{e:CS-coll-RG_K} \\
\frac{\partial}{\partial \log\mu_1}\, {}^{R}J(\bm{y}; \mu_i) &= - {}^{R\,}\gamma_J(\mu_1) \label{e:CS-coll-RG_J}
\end{align}
and similar expressions for the derivative with respect to $\mu_2$.

%%%%%%%%%%%%%%%%%%%%%%%%%%%%%%%%%%%%%%%%%%%%%%%%%%%%%%%%%%%%%%%%%%%%%%%%%%%%%%%
\section{Short distance matching}
\label{ss:matching}
%%%%%%%%%%%%%%%%%%%%%%%%%%%%%%%%%%%%%%%%%%%%%%%%%%%%%%%%%%%%%%%%%%%%%%%%%%%%%%%
Matching equations describe how the transverse momentum dependent and collinear DPDFs are related to each other by means of a convolution of the latter with a matching coefficient. In this section we describe the matching for short distances, with perturbative $|\tvec{z}_1|,\,|\tvec{z}_2| \ll |\tvec{y}|$, in which $\tvec{y}$ is fixed. For the DTMD/DPDF matching the relation is given by~\cite{Buffing:2017}
\begin{align}
& {}^{R}F_{a_1 a_2}(x_i,\tvec{z}_i,\tvec{y};\mu_i,\zeta) & = \sum_{b_1 b_2} 
	{}^{R\,}C\!_{a_1 b_1}(x_1',\tvec{z}_1;\mu_{1},\mu_{1}^2) \underset{x_1}{\otimes}
	{}^{R\,}C\!_{a_2 b_2}(x_2',\tvec{z}_2;\mu_{2},\mu_{2}^2) \underset{x_2}{\otimes}
	{}^{R}F_{b_1 b_2}(x_i',\tvec{y};\mu_{i},\zeta), \label{e:matching}
\end{align}
where the convolution between two functions $A$ and $B$ is given by
\begin{align}
A(x') \underset{x}{\otimes} B(x') &= \int_{x}^1 \frac{dx'}{x'}\, A(x')\,B\biggl(\frac{x}{x'}\biggr). \label{e:conv-def}
\end{align}
As can be seen, the matching consists of a convolution with two coefficient functions. These can largely be recycled from SPS. In Ref.~\cite{Buffing:2017} we have recalculated all coefficients at NLO level that were known in literature before~\cite{Aybat:2011zv,Bacchetta:2013pqa,Echevarria:2015uaa,Gutierrez-Reyes:2017glx}. We furthermore calculated the previously unknown $C_{\delta g\delta g}(x,\tvec{z})$, $C_{g\delta g}(x,\tvec{z})$ and $C_{q\delta g}(x,\tvec{z})$ at the same accuracy, which have not been calculated before since they do not contribute for SPS. In these coefficients $\delta g$ indicates linear gluon polarization.

The next step is to combine the evolution equations with the matching, giving rise to the combined evolution and matching equations. For the DTMDs, this gives us
\begin{align}
	\label{small-z-evolved}
& {}^{R}{F_{a_1 a_2}(x_i,\tvec{z}_i,\tvec{y};\mu_i,\zeta)}
\nonumber \\
&\quad = \exp\, \biggl\{ 
			\int_{\mu_{01}}^{\mu_1} \frac{d\mu}{\mu}\,
	\biggl[ \gamma_{F,a_1}(\mu, \mu^2)
		- \gamma_{K,a_1}(\mu) \log\frac{\sqrt{x_1\zeta/x_2}}{\mu} \biggr]
		+ {}^{R}{K}_{a_1}(\tvec{z}_1;\mu_{01})
		\log\frac{\sqrt{x_1\zeta/x_2}}{\mu_{01}}
\nonumber \\
& \hspace{13.25mm} + \int_{\mu_{02}}^{\mu_2} \frac{d\mu}{\mu}\,
	\biggl[ \gamma_{F,a_2}(\mu, \mu^2)
		- \gamma_{K,a_2}(\mu) \log\frac{\sqrt{x_2\zeta/x_1}}{\mu} \biggr]
		+ {}^{R}{K}_{a_2}(\tvec{z}_2;\mu_{02})
		\log\frac{\sqrt{x_2\zeta/x_1}}{\mu_{02}}
\nonumber \\
& \hspace{13.25mm} + {}^{R}{J}(\tvec{y};\mu_{0i})
		\log\frac{\sqrt{\zeta}}{\sqrt{\zeta_0}} \biggr\} \nonumber \\
&\quad \times\sum_{b_1 b_2} 
	{}^{R\,}{C}_{a_1 b_1}(x_1',\tvec{z}_1;\mu_{01},\mu_{01}^2)
	\underset{x_1}{\otimes}
	{}^{R\,}{C}_{a_2 b_2}(x_2',\tvec{z}_2;\mu_{02},\mu_{02}^2)
	\underset{x_2}{\otimes}
	{}^{R}{F_{b_1 b_2}(x_i',\tvec{y};\mu_{0i},\zeta_0)} \, .
\end{align}
Such an equation can be established at the cross section level as well, where the product of two DTMDs appears in the form
\begin{align}
	\label{W-large-y}
W_{\text{large $\tvec{y}$}} & = \sum_{R} \eta_{a_1 a_2}(R)\exp\, \biggl\{ 
			\int_{\mu_{01}}^{\mu_1} \frac{d\mu}{\mu}\,
	\biggl[ \gamma_{F,a_1}(\mu, \mu^2)
		- \gamma_{K,a_1}(\mu) \log\frac{{Q_1^2}}{\mu^2} \biggr]
		+ {}^{R}{K}_{a_1}(\tvec{z}_1;\mu_{01})
		\log\frac{{Q_1^2}}{\mu_{01}^2}
\nonumber \\
 & \hspace{12mm}\qquad\qquad + \int_{\mu_{02}}^{\mu_2} \frac{d\mu}{\mu}\,
	\biggl[ \gamma_{F,a_2}(\mu, \mu^2)
		- \gamma_{K,a_2}(\mu) \log\frac{{Q_2^2}}{\mu^2} \biggr]
		+ {}^{R}{K}_{a_2}(\tvec{z}_2;\mu_{02})
		\log\frac{{Q_2^2}}{\mu_{02}^2} \biggr\}
\nonumber \\[0.3em]
& \quad \times \sum_{c_1 c_2 d_1 d_2}
	{}^{R\,}{C}_{b_1 d_1}(\bar{x}_1',\tvec{z}_1;\mu_{01},\mu_{01}^2)
		\underset{\bar{x}_1}{\otimes}
	{}^{R\,}{C}_{b_2 d_2}(\bar{x}_2',\tvec{z}_2;\mu_{02},\mu_{02}^2)
\nonumber \\[0.3em]
& \hspace{10.8mm}\quad \underset{\bar{x}_2}{\otimes}
	{}^{R\,}{C}_{a_1 c_1}(x_1',\tvec{z}_1;\mu_{01},\mu_{01}^2)
		\underset{x_1}{\otimes}
	{}^{R\,}{C}_{a_2 c_2}(x_2',\tvec{z}_2;\mu_{02},\mu_{02}^2)
\nonumber \\
& \quad \underset{x_2}{\otimes} \bigl[ \Phi(\nu \tvec{y}) \bigr]^2\,
	\exp\,\biggl[ {}^{R}{J}(\tvec{y};\mu_{0i})
				\log\frac{Q_1 Q_2}{\zeta_0} \,\biggr]
	{}^{R}{F_{d_1 d_2}(\bar{x}_i,\tvec{y};\mu_{0i},\zeta_0)}\,
	{}^{R}{F_{c_1 c_2}(x_i,\tvec{y};\mu_{0i},\zeta_0)} \, ,
\end{align}
where $\eta_{a_1 a_2}(R)$ is a sign factor explained in detail in~\cite{Buffing:2017}. At the level of the cross section, there are four coefficient functions, two for each DPDF. In the above equation, $\Phi(\nu \tvec{y})$ is a function needed to regulate ultraviolet divergences \cite{Diehl:2017kgu}.

It is important to stress that in the equation for the cross section no rapidity parameters $\zeta$'s are present anymore, since they can be traded for the energy scales $Q_1^2$ and $Q_2^2$ through the relation $\zeta\overline{\zeta}=Q_{1}^2 Q_{2}^2$. An important difference with the single TMD/PDF matching is the presence of an additional Sudakov suppression $\exp{\left[ {}^{R}{J}(\tvec{y};\mu_{0i})\log\frac{Q_1 Q_2}{\zeta_0}\right]}$. In the color singlet channel ($R=1$), we have ${}^{1}J=0$ and this term disappears.

%%%%%%%%%%%%%%%%%%%%%%%%%%%%%%%%%%%%%%%%%%%%%%%%%%%%%%%%%%%%%%%%%%%%%%%%%%%%%%%
\section{Discussions and conclusions}
\label{ss:conclusions}
%%%%%%%%%%%%%%%%%%%%%%%%%%%%%%%%%%%%%%%%%%%%%%%%%%%%%%%%%%%%%%%%%%%%%%%%%%%%%%%
In these proceedings we have looked at double parton scattering (DPS). Using an approach where both the partonic processes are in the perturbative regime and well separated from each other in configuration space, we have given the soft function and evolution equation for DPS, both of which derived in \cite{Buffing:2017}. We then proceeded to give the evolution equations for both the rapidity and energy scale evolution of DTMDs, the solutions of which we combined into an equation describing the evolution with respect to both rapidity and energy scales. Although certain aspects of the solutions are equivalent to doubling the SPS results, there are further contributions in the DPS situation, since an additional $\tvec{y}$-dependent Sudakov suppression for the color non-singlet configuration appears. A determination of the size of such suppressions will be the topic of future work.

Subsequently, we have looked at the DTMD/DPDF matching equations. The DTMDs are related to the DPDFs by a matching with two coefficient functions, one for each parton. These coefficients can largely be recycled from SPS literature~\cite{Aybat:2011zv,Bacchetta:2013pqa,Echevarria:2015uaa,Gutierrez-Reyes:2017glx}, although a few coefficients are new, since they are not needed for SPS. Therefore, Ref.~\cite{Buffing:2017} is the first place where $C_{\delta g\delta g}(x,\tvec{z})$, $C_{g\delta g}(x,\tvec{z})$ and $C_{q\delta g}(x,\tvec{z})$ will be given. Combining the matching equations with the solutions of the evolution equations, we have obtained expressions for combined matching/evolution of DTMDs and a corresponding expression for the $p_T$-resummed cross section.
%at both the level of DTMDs as for cross section contributions.

%%%%%%%%%%%%%%%%%%%%%%%%%%%%%%%%%%%%%%%%%%%%%%%%%%%%%%%%%%%%%%%%%%%%%%%%%%%%%%%

%%%%%%%%%%%%%%%%%%%%%%%%%%%%%%%%%%%%%%%%%%%%%%%%%%%%%%%%%%%%%%%%%%%%%%%%%%%%%%%

\end{document}